\documentclass[aps,prl,twocolumn,superscriptaddress]{revtex4-2}
\usepackage[utf8]{inputenc}
\usepackage{graphicx}
\usepackage{amsmath}
\usepackage{amssymb}

\usepackage{color}
\definecolor{blue(pigment)}{rgb}{0.2, 0.2, 0.6}
\definecolor{red(pigment)}{rgb}{0.6, 0.2, 0.2}
\definecolor{blue2(pigment)}{rgb}{0.0, 0.1, 0.9}

\begin{document}

\title{
A minimal phase-coupling model for intermittency in turbulent systems
}
\noaffiliation
\author{Jos\'e-Agust\'in Arguedas-Leiva}
\affiliation{Max Planck Institute for Dynamics and Self-Organization, Am Fa\ss berg 17, 37077 G\"ottingen, Germany}
\author{Enda Carroll}
\affiliation{School of Mathematics and Statistics, University College Dublin, Belfield, Dublin 4, Ireland}
\author{Luca Biferale}
\affiliation{Dept. Physics and INFN, University of Rome Tor Vergata, Via Ricerca Scientifica 1, 00133 Rome, Italy}
\author{Michael Wilczek}
\email{michael.wilczek@ds.mpg.de}
\affiliation{Max Planck Institute for Dynamics and Self-Organization, Am Fa\ss berg 17, 37077 G\"ottingen, Germany}
\affiliation{Theoretical Physics I, University of Bayreuth, Universit\"atsstr.~30, 95447 Bayreuth, Germany}
\author{Miguel D. Bustamante}
\affiliation{School of Mathematics and Statistics, University College Dublin, Belfield, Dublin 4, Ireland}
\date{\today}

\begin{abstract}
Turbulent systems exhibit a remarkable multi-scale complexity, in which spatial structures induce scale-dependent statistics with strong departures from Gaussianity. In Fourier space, this is reflected by pronounced phase synchronization. A quantitative relation between real-space structure, statistics, and phase synchronization is currently missing. Here, we address this problem in the framework of a minimal deterministic phase-coupling model, which enables a detailed investigation by means of dynamical systems theory and multi-scale high-resolution simulations. We identify the spectral power-law steepness, which controls the phase coupling, as the control parameter for tuning the non-Gaussian properties of the system. Whereas both very steep and very shallow spectra exhibit close-to-Gaussian statistics, the strongest departures are observed for intermediate slopes comparable to the ones in hydrodynamic and Burgers turbulence. We show that the non-Gaussian regime of the model coincides with a collapse of the dynamical system to a lower-dimensional attractor and the emergence of phase synchronization, thereby establishing a dynamical-systems perspective on turbulent intermittency.
\end{abstract}

\maketitle

%%%%%%%%%%%%%%%%%%%%%%%%%%%%%%%%%
\begin{figure}[t]
\includegraphics[width=\columnwidth]{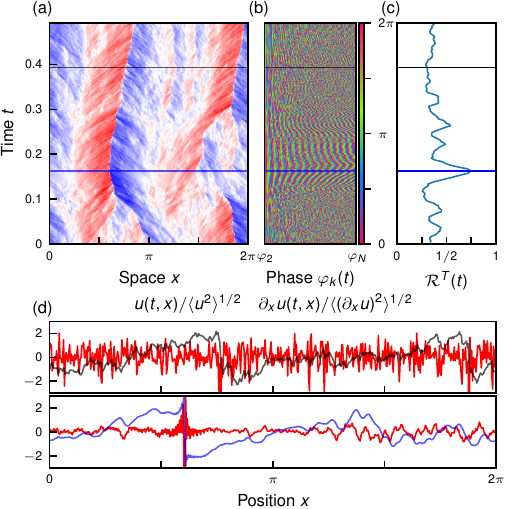}
\caption{Numerical simulations of \eqref{eq:oscillator_system} with $k_0=1$ and $N=2^{9}$, with a Burgers-like steepness parameter $\alpha=1$. (a) Plot of the real-space field $u(t, x)$ displaying a shock near $x=\pi/2$ at the time indicated by the blue line. The field $u(t, x)$ is obtained by solving the phase dynamics \eqref{eq:oscillator_system} and inserting the time-evolving phases into \eqref{eq:RealSpaceu} for prescribed amplitudes $a_k$. (b) Plot of individual phases $\varphi_2, \ldots \varphi_N$. The gray line marks an instance of a relatively disordered regime while the blue line marks a relatively synchronized regime. (c) Time-dependent order parameter $\mathcal{R}^T(\alpha,t)$, cf.~Eq.~\eqref{eq:R_avg}, for the synchronization of the system (here with $T=0.0244$). The peak corresponds to a synchronization event related to a real-space shock. (d) Snapshots of the real-space field $u(t,x)$ in the disordered (top, gray) and synchronized (bottom, blue) regimes. Red curves show the gradient $\partial_x u (t, x)$ to illustrate the difference between the two regimes.}
\label{fig:figure_1}
\end{figure}
\begin{figure}
    \includegraphics[width=0.5\textwidth]{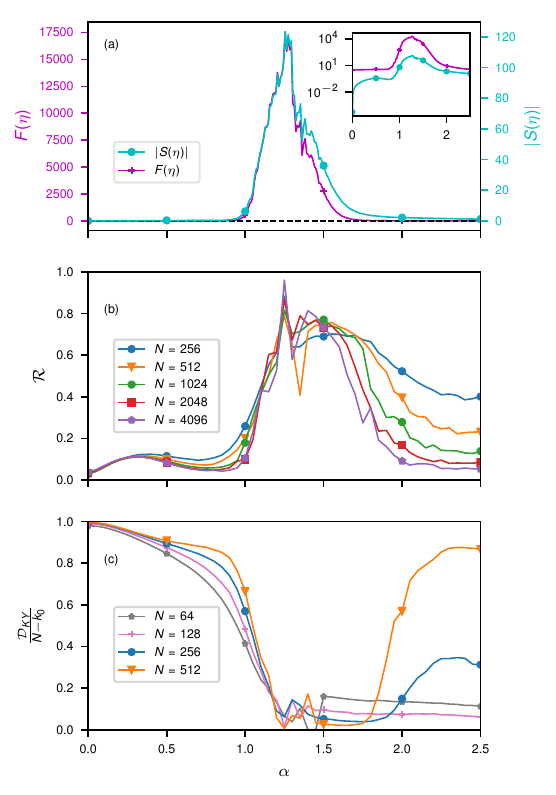}
    \caption{(a): Absolute value of the skewness  $|S(\eta)|$ and the flatness $F(\eta)$ for the smallest increment $\eta$ as a function of $\alpha$ (inset: same figure on log-lin scales, $N=2^{15}$). (b): the time and scale-averaged phase synchronization $\mathcal{R}(\alpha)$ as a function of $\alpha$ for various system sizes. (c): ratio between  the Kaplan-Yorke dimension $\mathcal{D}_{KY}$ and $N - k_0$ as a function of $\alpha$ for various system sizes.}
    \label{fig:figure_3}
\end{figure}

%%%%%%%%%%%%%%%%%%%%%%%%%%%%%%%%%
{\sc Introduction.} Turbulence is a prototypical non-equilibrium phenomenon with a large number of strongly interacting degrees of freedom \cite{Frisch95,sreenivasan1999fluid,Pope00,davidson2015turbulence,alexakis2018cascades,biskamp2003magnetohydrodynamic}, exhibiting strong departures from Gaussianity on the smallest spatial scales. In real space, non-Gaussian fluctuations can be related to coherent, intense, and rare  events in the velocity  gradients -- a phenomenon also dubbed as {\it intermittency} \cite{parisi1985multifractal,kraichnan1990models}. Intermittency  can also be studied from the complementary perspective of Fourier space. While Gaussian random fields feature completely uncorrelated phases, phase correlations can give rise to complex scale-dependent properties, as the ones developed in the presence of coherent shocks.
 Elucidating these connections is important for both fundamental and applied aspects. In particular, we currently miss a clear identification of which dynamical degrees of freedom lead to such bursting and quiescent chaotic alternations of temporal and spatial flow realizations. As a result, we lack optimal protocols to avoid disrupting fluctuations in engineering turbulence \cite{jimenez2012cascades,duriez2017machine}, predict extreme events in geophysical flows \cite{ragone2018computation,BiferalePRX2016} and  control existence and uniqueness of the PDE solutions \cite{fefferman2006existence}, just to cite a few open problems with multidisciplinary impacts. Studying these issues in fully developed three-dimensional turbulence is an extremely challenging task. The hope is to isolate the main aspects of this problem in simpler, more tractable models. One popular way is to lower the complexity by mode reduction, as in the case of sub-grid-scale modeling \cite{meneveauARFM,stevens_wilczek_meneveau_2014}, Fourier surgery \cite{biferale2003shell,bohr2005dynamical}, statistical closure \cite{lesieur2012turbulence}, partial freezing of some spectral degrees of freedom \cite{shejackson,biferaleprl2019}  or asymptotic expansions \cite{zakharov2012kolmogorov,nazarenko2011wave}. All attempts have merits and deficiencies, the main common drawback being the  compromised ability to describe simultaneously spatial and temporal fluctuations on a wide range of scales. Notably, only very few studies have addressed the connection between the emergence of coherent intermittent structures in real space and phase correlations, connecting the presence of bursts of spectral energy fluxes (and dissipation) with  Fourier phase dynamics \cite{buzzicotti2016phase, reynolds2016fourier, wilczek2017emergence, mastersthesis_arguedas_2017, MurrayBustamante2018}.
 \newline
In this letter we combine theory and simulations to provide a dynamical systems link between real-space intermittency and phase correlations in Fourier space. We do so by means of a minimal deterministic description of hydrodynamic turbulence derived from a PDE model, preserving the whole richness of multi-scale spatial and temporal statistics. The model is formulated in terms of Fourier phases whose dynamical coupling resembles the one in Navier-Stokes turbulence: specifically, it is Burgers turbulence with the important distinction that the amplitudes are kept at fixed values such that only the phases evolve, obeying a deterministic system that supports a turbulent attractor.\newline
By changing the energy spectrum slope we can tune the coupling strength of the phases and study how the dynamics (intermittency) changes.
We find that the system transitions to non-Gaussian statistics as the spectrum is gradually steepened. For slopes beyond a certain value, the rare fluctuations become less extreme and return to near-Gaussian statistics. Strikingly, the strongest deviations occur in the intermediate range, within the range of values attained by turbulent systems. Within this range, the dimension of the strange attractor  collapses to a minimum, indicating that non-Gaussian real-space statistics are related to the collapse of the dynamical system onto a lower-dimensional manifold.\newline
Our analysis sheds light on the emergence of coherent structures and the associated phase synchronization phenomena \footnote{In this paper, \emph{synchronization} is understood as a transient state whereby the phases of the Fourier modes over an extended range of spatial scales evolve following similar patterns, showing strong correlations during finite time intervals. This is  similar to the definition used in classical phase models \cite{rosenblum1996phase, boccaletti2002synchronization}, although our work is the first, to our knowledge, to study synchronization in a Fourier phase model based on a system with quadratic nonlinearity.}, establishing connections between the statistical theory of non-equilibrium systems and dynamical systems theory.\newline
%%%%%%%%%%%%%%%%%%%%%%%%%%%%%%%%%
{\sc The model.} As a starting point, let us consider the one-dimensional Burgers equation
\begin{equation}
    \partial_t u(t,x)+u(t,x)\partial_x u(t,x)=\nu \partial_x^2 u(t,x).
    \label{eq:burgers}
\end{equation}
This simple prototypical PDE is reminiscent of the Navier-Stokes equations, known to develop multi-scale bifractal scaling properties, shocks, non-Gaussian statistics and many other non-trivial statistical features \cite{zikanov1997statistics,balkovsky1997intermittency,bec2007burgers,da1994stochastic,chekhlov1995kolmogorov,eyink2015spontaneous,she1992inviscid,bouchaud1995scaling,weinan2000invariant,buzzicotti2016intermittency}.  
We consider a one-dimensional field $u(t,x)$ on a $2\pi$-periodic domain with Fourier decomposition
\begin{equation}
u(t,x)=\underset{k\in \mathbb{Z}}{\sum}a_k(t)\: \exp\left(\mathrm{i}\left[(\varphi_k(t) + kx\right]\right).
\label{eq:RealSpaceu}
\end{equation}
By inserting \eqref{eq:RealSpaceu} into \eqref{eq:burgers}, we obtain the evolution for the amplitudes and the phases
\begin{align}
a_k\dfrac{\mathrm{d}\varphi_k}{\mathrm{d}t}=&-\dfrac{k}{2}\underset{p\in \mathbb{Z}}{\sum}a_p\: a_{k-p}\: \cos(\varphi_p+\varphi_{k-p}-\varphi_k), \label{eq:burgers_phases}\\
\dfrac{\mathrm{d}a_k}{\mathrm{d}t}=&\dfrac{k}{2}\underset{p\in \mathbb{Z}}{\sum}a_p\: a_{k-p}\: \sin(\varphi_p+\varphi_{k-p}-\varphi_k)-\nu k^2 a_k . \label{eq:burgers_amplitudes}
\end{align}
This infinite set of coupled ODEs describes the full Burgers dynamics of the Fourier phases and amplitudes. 
Recently, we showed that the dynamics of the Fourier phases $\varphi_k(t)$ determine to a great extent the shock dynamics and the associated non-Gaussian statistics when the amplitudes follow a Burgers-like scaling \cite{MurrayBustamante2018}. In order to establish a deeper dynamical systems understanding of the role of Fourier phases in turbulence, we now follow a different approach, by introducing a minimal {\it deterministic} model enjoying a turbulent attractor. Take equation \eqref{eq:burgers_phases}  as a starting point and  set the amplitudes to prescribed constants \begin{equation}
\label{eq:ak}
    a_k = \vert k\vert^{-\alpha}, \,\, |k| > k_0, \qquad a_k = 0, \,\, |k| \leq k_0,
\end{equation}
where the steepness $\alpha$ is our new  {\it continuous control parameter} and  $k_0 > 0$ is a large-scale cutoff leading to a finite integral length scale, which destabilizes a single-shock-like fixed point, allowing thus for %non-steady dynamics and 
a turbulent attractor.
The phase dynamics is obtained from  \eqref{eq:burgers_phases} which becomes a system of coupled oscillators $\varphi_k$ satisfying
\begin{equation}
\dfrac{\mathrm{d}\varphi_k}{\mathrm{d}t}=\underset{p\in \mathbb{Z}}{\sum}\omega_{k,p}\;\cos(\varphi_p+\varphi_{k-p}-\varphi_k), \quad |k| > k_0\,,
\label{eq:oscillator_system}
\end{equation}
with coefficients $\omega_{k,p} = -k\; |p(k-p)|^{-\alpha}\; |k|^{\alpha}$ when $|k-p|, |p|>k_0$ ($\omega_{k,p}=0$ otherwise), and with $\varphi_{-k} = -\varphi_k$ (reality condition). Compared to equation \eqref{eq:burgers_phases}, we have rescaled time in \eqref{eq:oscillator_system} to absorb the factor $1/2$. The triadic interaction term couples the phases with wavenumbers $k$, $p$, and $k-p$, via the so-called  triad phase $\varphi_{p, k-p}^k := \varphi_p+\varphi_{k-p}-\varphi_k$. It is important to note that this phase-only model  does not need an energy input/output mechanism, as constant energy is maintained by the constant amplitudes. Furthermore, it is formally fully time reversible under the symmetry $ t \to -t;  \varphi_k \to \varphi_k + \pi$. However, it will not come as a surprise that, like in a formally reversible version of the Navier-Stokes equations  \cite{Ga000a,Ga006c,BCDGL018,SDNKT018}, the chaotic dynamics spontaneously break the time symmetry leading to a non-Gaussian and skewed velocity increment probability density function (PDF).
To study the model numerically, we further introduce a discretization with grid spacing $\Delta x = \pi/N$, effectively setting $a_k = 0, |k|> N$. The reality condition $\varphi_{-k} = -\varphi_k$ leaves us with a set of phases evolving on modes $ k_0 < k \leq N$. We set $k_0=1$ so $a_1 = a_{-1} = 0$ and thus the evolving variables are $\varphi_2, \ldots, \varphi_{N}$. 
Note that the energy spectrum of the field is fixed and perfectly self-similar: $E_k \sim a_k^2$, with a power-law decay of $E_k\propto k^{-2\alpha}$. The observed original Burgers case, where quasi discontinuities (shocks) dominate the high-order statistics,  corresponds to $\alpha=1$.\newline
%%%%%%%%%%%%%%%%%%%%%%%%%%%%%%%%%
{\sc Numerical results on real-space and phase dynamics \& statistics.} We integrate numerically \eqref{eq:oscillator_system} with a fourth-order Runge-Kutta method starting from uniformly random initial conditions. The nonlinear term can be written as a convolution, which we efficiently evaluate with a pseudospectral method.
Figure \ref{fig:figure_1} illustrates the dynamics of our model, for the choice of steepness $\alpha=1$ (Burgers case), revealing insights into the relation between non-Gaussianity of the real-space statistics and Fourier phase synchronization. Panel (a) is a space-time plot of the velocity field from this minimal model, showing that shocks are the dominant structure. As time evolves, shocks steadily merge and separate. Occasionally, they merge into one dominating shock (horizontal blue line). Panel (b) shows a time series of the individual Fourier phases of the model. It shows that the presence of this dominating shock corresponds to highly ordered patterns in the phase plot. Away from these events, the system is dominated by smaller shocks and we observe a low coherence (gray line). To quantify this, panel (c) shows the time average phase synchronization \eqref{eq:R_avg}, which  reveals the synchronization of the oscillator system locally in time. The time of the highest synchronization corresponds to the dominating shock in real space. Panel (d) shows that the %synchronization events and 
presence of the dominating shock (blue line) yield extreme events in the gradient field characterizing the small scales of the velocity field.\newline
%%%%%%%%%%%%%%%%%%%%%%%%%%%%%%%%%
\begin{figure*}
\includegraphics[width=\textwidth]{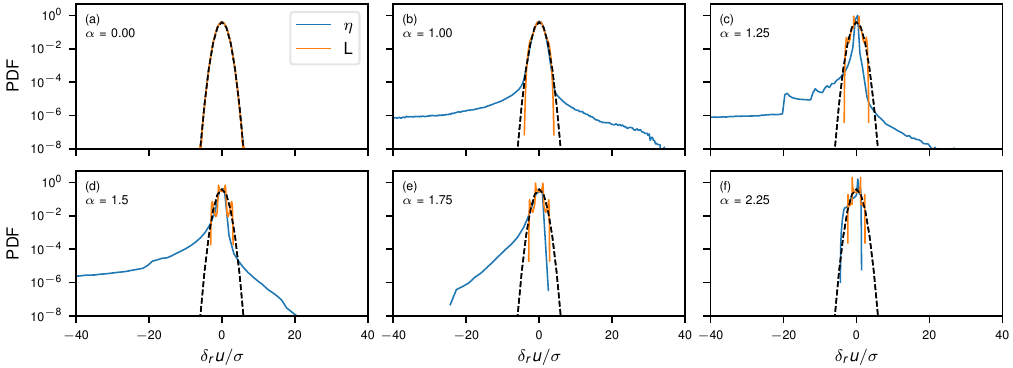}
\caption{
(a)-(f): Standardized probability density distributions (PDFs) of $\delta_r u$ calculated at the smallest, $\eta$, and largest increments, $L$. (a) For flat Fourier amplitudes $\alpha=0$ the velocity field is Gaussian across all scales. (b)-(e) Increasing $\alpha$ leads to heavy tails at small scales. (f) For a steep enough spectrum the velocity field is dominated by the first few modes leading to PDFs without heavy tails. Data for $N=2^{15}$ and $k_0=1$.}
\label{fig:figure_2}
\end{figure*}
%%%%%%%%%%%%%%%%%%%%%%%%%%%%%%%%%
By changing the free parameter $\alpha$ in \eqref{eq:oscillator_system} we  control the multi-scale  coupling among the phases and the hierarchical organization of typical time scales. In a {\it local} approximation, i.e.~supposing the dynamics at wavenumber $k$ is mainly driven by triads around the same wavenumber, $|k| \sim |k-p| \sim |p|$, we can estimate the scale-dependent eddy-turnover time as $\tau_k \sim |k|^{\alpha-1}$, indicating that, within this approximation, we reach a regime where small scales are faster than the large  scales if $\alpha<1$ (and slower if $\alpha>1$). The local-triad approximation is expected to be valid in the range $0.5 < \alpha < 1.5$ \cite{rose1978fully}, where the Fourier transform connecting the spectrum and the two-point velocity correlation function does not diverge neither in the UV nor in the IR.  As a result, we expect that in the above range and around $\alpha=1$ a non-trivial balancing between spatial and temporal fluctuations will set in.\newline
In figure \ref{fig:figure_2} we indeed observe that our model has non-trivial scale- and steepness-dependent statistics. Here we show the probability distribution functions (PDF) of the velocity increments $\delta_r u = u(x+r)-u(x)$ for two different scales, $r= L := \pi$ and $r = \eta := \pi/N$, denoting the largest and smallest distances in the periodic domain, respectively. Real-space statistics are obtained by inserting the phase dynamics into \eqref{eq:RealSpaceu}. 
For completely uniform amplitudes (steepness $\alpha=0$) the phases evolve under an all-to-all coupling with equal strength. Note that this choice of spectral amplitudes corresponds to a delta-correlated field in real space. In this case, all phases become dynamically uniformly distributed and uncorrelated, leading to a Gaussian velocity field at all scales (panel (a) in figure \ref{fig:figure_2}). In contrast, for  steepness values within the range $[0.5, 1.5]$, where the local-triad approximation is expected to be valid, heavy tails are observed in the velocity increment PDF at small scales (panels (b)--(d) in figure \ref{fig:figure_2}). For the smallest increment, the negative PDF tails are much heavier than the positive tails and both are much heavier than Gaussian. Heuristically (to be quantified later), this is the result of phase synchronization leading to shocks (anti-shocks), i.e.~extreme negative (positive) gradients.\newline
The presence of extreme events is maximal at  $\alpha\sim 1.25$, as evidenced in figure \ref{fig:figure_2}(c). For higher values of $\alpha$ the PDF tails slowly regularize. In this limit, the large-scale modes dominate the real-space velocity field, leading to  a dominant sinusoidal mode with superimposed smaller fluctuations. As a consequence, the large $\alpha$ limit shows close-to-Gaussian statistics throughout.\newline
To quantify the steepness-dependent departure of the small scales from Gaussianity we measure the skewness and flatness: 
\begin{align}
S(r) = \frac{\langle (\delta_r u)^3\rangle}{\langle (\delta_r u)^2\rangle^{3/2}} \,,\qquad  F(r) = \frac{\langle (\delta_r u)^4\rangle}{\langle (\delta_r u)^2\rangle^2} \, .
\label{eq:flatness_skewness}
\end{align} 
%Note that d
Due to our frozen-amplitude condition the denominators of both quantities do not fluctuate. Figure \ref{fig:figure_3}(a) shows a clear transition at $\alpha \sim 1.0$. The peaks of skewness and flatness at $\alpha \sim 1.25$ correspond to the presence of extremely intense negative gradients seen in figure \ref{fig:figure_2}(c).\newline
As the steepness is increased further, the phases evolve under a non-local and non-trivial triad coupling. This gives rise to synchronization events, which underlie the steepness-dependent transition observed in the real-space statistics. Note, however, that when the steepness is too large the timescales from the triad coupling can get too separated, as the coefficients $\omega_{k,p}$ in (\ref{eq:oscillator_system}) become too small when $|p|$ and $|k-p|$ are large. Thus we expect to see synchronization over a finite range of steepness values only. In the next sections we will quantify the dependence on the $\alpha$ parameter, of synchronization, and of the structure of the associated chaotic attractors.\newline
{\sc Synchronization.} We quantify the  behaviour of triad phases across a range of scales for  (\ref{eq:oscillator_system}) by defining the scale-dependent collective phase $\theta_k$: 
\begin{align}
    e^{\mathrm{i} \theta_k} = \frac{\sum_{p \in \mathbb{Z}} a_{p}a_{k - p} e^{\mathrm{i} (\varphi_p + \varphi_{k - p} - \varphi_k)} } {\left|\sum_{p\in \mathbb{Z}} a_p a_{k - p}e^{\mathrm{i} (\varphi_p + \varphi_{k - p} - \varphi_k)}\right| }.
    %\theta_k = \arg \big( \sum_{p} a_{p}a_{k - p} e^{i (\varphi_p + \varphi_{k - p} - \varphi_k)} \big).
\end{align}
This collective phase is dynamically relevant as the RHS of  \eqref{eq:oscillator_system} is proportional to $\cos\theta_k$. The fluctuations of $\theta_k$ over time serve as a measure of the triad phase coherence across scales. Thus, averaging over a causal time window $T$ from $t-T$ to $t$, we get the following scale-dependent Kuramoto order parameter:
\begin{align}
    R^T_k(t) e^{\mathrm{i} \Theta^T_k(t)} = \left\langle e^{\mathrm{i}\theta_k(t)}\right\rangle_T
    \label{eq:scale_dep_order_parameter}
\end{align}
As usual we have $0\leq R^T_k \leq 1$, and phase synchronization is indicated by $R^T_k$ values close to $1$.  Averaging additionally over the spatial scales, we define the average phase synchronization by
\begin{align}
\label{eq:R_avg}
    \mathcal{R}^T(\alpha,t) = \frac{1}{N-k_0}\sum_{k=k_0+1}^{N} R^T_k(t),
\end{align}
which measures how the phase synchronization changes as a function of the spectral slope. As discussed earlier, we evaluated the time-dependent average phase synchronization $\mathcal{R}^T$ in Fig.~\ref{fig:figure_1}(c) to establish the correspondence between real-space structures and phase synchronization. For very large $T$, we obtain the time and scale-averaged phase synchronization $\mathcal{R}(\alpha) = \lim_{T \rightarrow\infty}  \mathcal{R}^T(\alpha,t)$.

Figure \ref{fig:figure_3}(b) shows the average phase synchronization $\mathcal{R}(\alpha)$ as a function of $\alpha$ for various system sizes $N$. The relatively high synchronization  seen for small $N$ at $\alpha > 2.0$ decreases as $N$ is increased. This is due to the addition of faster and noisier oscillators to the system causing a convergence towards a pronounced peak for $\alpha \in [1.0, 2.0]$, indicating high phase synchronization for this interval for large $N$. The synchronization peak is remarkably coincidental with the flatness and skewness peaks shown in figure \ref{fig:figure_3}(a), providing quantitative evidence in support of the relation between synchronization (a dynamical-system measure) and intermittency (a real-space measure).
%%%%%%%%%%%%%%%%%%%%%%%%%%%%%%%%%
%%%%%%%%%%%%%%%%%%%%%%%%%%%%%%%%%

\noindent {\sc Chaos characterization.} 
As an additional characterization of the dynamical system, we estimate the properties of the underlying strange attractor as a function of $\alpha$ and for $N=64, 128, 256, 512$ by examining the Lyapunov exponents (LEs)  \cite{Ginelli2007a}.
 For reasons of numerical complexity we cannot reach the same resolution we used for the statistical characterization of intermittency and synchronization; however as we will see below the $N=512$ case shows strong indications of convergence to the large-$N$ limit.\newline
Using the LEs we can calculate the dimension of the attractor via the Kaplan-Yorke (KY) {approximation} \cite{kaplan1979chaotic, frederickson1983liapunov}. Given the ordered LEs $\lambda_1 \geq \lambda_2 \geq \dots \geq \lambda_{N - k_0}$,
the KY dimension is defined as
\begin{equation}
    \mathcal{D}_{KY} = J + \frac{\sum_{j = 1}^{J}\lambda_j}{|\lambda_{J + 1}|} \,,
    \label{eq:kaplan_yorke}
\end{equation}
where the conditions $\sum_{j=1}^{J} \lambda_j\geq 0$ and $\sum_{j = 1}^{J + 1} \lambda_j < 0$ define the index $J$. The KY dimension gives a measure of the systems' effective degrees of freedom. Figure \ref{fig:figure_3}(c) shows a plot of the ratio between the $\mathcal{D}_{KY}$ and the number of available degrees of freedom, as a function of $\alpha$ and for several values of  $N$. It is evident that as $N$ grows a clear pattern emerges, whereby $\mathcal{D}_{KY}$ greatly diminishes for values of $\alpha$ inside the interval $[1.0, 2.0]$, a behaviour that coincides, on the one hand, with the departure from Gaussianity observed in figure \ref{fig:figure_3}(a), and on the other hand, with the increase in phase synchronization shown in figure \ref{fig:figure_3}(b).\newline
%%%%%%%%%%%%%%%%%%%%%%%%%%%%%%%%%
{\sc Conclusions.} Our minimal model sheds light into the nature of coherent structures as low-dimensional objects, establishing a dynamical scenario where real-space intermittency and phase synchronization are accompanied by a reduction in the dimensions of the attractor. 
In our model coherent structures are controlled by Fourier phase dynamics only, as the energy spectrum is static and plays a background role. %\newline
Our results open new perspectives concerning the possibility to connect turbulence intermittency with dynamical system tools based on phase synchronization and chimera states \cite{shepelev2018chimera}.\newline
On the quantitative side, our results provide insight into the solution to the full inviscid Burgers equation, where all amplitudes are allowed to evolve. There, for generic initial conditions, a finite-time singularity develops characterized by phase synchronization and a power-law spectrum with steepness $1.33 \leq \alpha \leq 1.50$ \cite{SULEM1983138, PhysRevE.86.066302}. We have checked that this  behaviour is robust, occurring even under the constraint $a_{k_0}=0$ for $k_0=1$. Because in the full equations the spectrum evolves slowly, it is natural to expect that in our frozen-spectrum constrained model the phases must show high correlation in the same range of imposed slopes.\newline
A natural extension of this work would be an investigation of the phase-only 3D Navier-Stokes dynamics by fixing the amplitudes of all Fourier modes, including comparisons to Navier-Stokes equations with a fixed spectrum, either for all wavenumbers or for a subset of them \cite{shejackson,biferaleprl2019}. Results in this direction would help to shed additional light on the origin of extreme events and small-scale intermittency. \newline
Acknowledgment. This work received funding from the European Research Council (ERC)
under the European Union’s Horizon 2020 research and innovation
programme (grant agreement No 882340). JAAL and MW were supported by the Max Planck Society. EC and MDB were supported by the Irish Research Council under grant number GOIPG/2018/2653.
%%%%%%%%%%%%%%%%%%%%%%%%%%%%%%%%%
\bibliography{references}
\end{document}